# Relationship between low-discrepancy sequence and static solution to multi-bodies problem

Feng Wu[*], Yuelin Zhao, Ke Zhao, and Wanxie Zhong

*State Key Laboratory of Structural Analysis of Industrial Equipment,*

*Department of Engineering Mechanics, Faculty of Vehicle Engineering and*

*Mechanics, Dalian University of Technology, Dalian 116023, P.R.China*

*Email: wufeng_chn@163.com, zhaoyl9811@163.com, zhaoke_93@163.com,*

*zwoffice@dlut.edu.cn*

[*]Corresponding author: Tel: 8613940846142; E-mail address: wufeng_chn@163.com

October 2021

**Abstract:** The main interest of this paper is to study the relationship between the low-discrepancy sequence and the static solution to the multi-bodies problem in high-dimensional space. An assumption that the static solution to the multi-bodies problem is a low-discrepancy sequence is proposed. Considering the static solution to the multi-bodies problem corresponds to the minimum potential energy principle, we further assume that the distribution of the bodies is the most uniform when the potential energy is the smallest. To verify the proposed assumptions, a dynamical evolutionary model (DEM) based on the minimum potential energy is established to find out the static solution. The central difference algorithm is adopted to solve the DEM and an evolutionary iterative scheme is developed. The selection of the mass and the damping coefficient to ensure the convergence of the evolutionary iteration is discussed in detail. Based on the DEM, the relationship between the potential energy and the discrepancy during the evolutionary iteration process is studied. It is found that there is a significant positive correlation between them, which confirms the proposed assumptions. We also combine the DEM with the restarting technique to generate a series of low-discrepancy sequences. These sequences are unbiased and perform better than other low-discrepancy sequences in terms of the discrepancy, the potential energy, integrating eight test functions and computing the statistical moments for two practical stochastic problems. Numerical examples also show that the DEM can not only generate uniformly distributed sequences in cubes, but also in non-cubes.

# 1. Introduction

Monte Carlo method (MCM) is an important statistical method, which is mainly used to solve high-dimensional stochastic problems. In recent years, it has played an important role in many fields, such as structural safety and reliability analysis [1,2], quantum physics [3,4], finance [5,6], network [7], artificial intelligence [8], etc . In the MCM, the random outputs of all the samples in the given random sampling sequence should be calculated first, and then these random outputs are used to evaluate the statistical results. Hence, the accuracy of the MCM mainly depends on the number of samples. The convergence speed of the MCM is of $O(n^{-0.5})$, which means a very large number of samples and a large amount of calculation cost are required [9]. It has been found that the accuracy of the MCM is closely related to the discrepancy of the sample sequence. The discrepancy of the sample sequence refers to the difference between the empirical distribution evaluated based on this sample sequence and the real probability distribution. Discrepancy can be used to measure the degree of uniformity of the sequence distribution. The smaller the discrepancy, the better the uniformity of the sequence and the better the accuracy of the MCM results based on the sequence. Therefore, many scholars were engaged in the research of sequence discrepancy, and constructed a series of low-discrepancy sequences. The MCM based on low-discrepancy sequence is generally called quasi-Monte Carlo method (QMCM).

There have been existing many types of low-discrepancy sequences, such as Hua Wang sequence, Halton sequence, Kronecker sequence, van der Corput point sequence, Niederreiter sequence, Soble sequence, and so on [10-15], and all of these sequences are based on the number theory. However, a major disadvantage of these sequences is the strong correlation between points in higher dimensions, which is the low-dimensional projections of the points frequently lie close together [10-15]. There are some ways to improve this problem now. One method is to randomize the sequence, such as the random scrambling method. One advantage of this method is that the randomize sequences are random, and hence the Monte Carlo theory can be used to evaluate the error. The randomized technique can eliminate the correlation of the sequence. Another method is to use some heuristic optimization algorithms (e.g., the genetic algorithm) to optimize the existing sequence [16-18]. This kind of method takes the discrepancy (e.g., the star discrepancy) as the objective function, and uses the evolutionary method to reduce the discrepancy of the sequence. Previous studies have shown that the uniformity of the sequence generated by using the optimization method is often better than that generated by the first kind of method [16-18], which means how to optimize the sequence efficiently is a very valuable topic.

The multi-bodies problem is a classical problem in physics, which is also one of the most difficult problems [19,20] in physics. For a long time, scientists have paid great attention to finding the dynamic solutions to the multi-bodies problems [21-25], and the static solution to multi-bodies problem is rarely studied. It seems that the static solution is not very useful. Actually, there are many simple atomic crystals in nature, and their atoms are arranged very regularly and evenly. Imagine that *n* particles with the same mass are distributed in a cube space, there is gravitation between each particle. Under the action of gravitational field, the particles finally reach the state of force equilibrium. Is the distribution of particles uniform? Take the two-dimensional graphene as an example. The interaction force between each carbon atom is the same, and the carbon atoms at the state of force equilibrium are evenly arranged in a two-dimensional plane in the form of regular hexagon. In fact, there are many homogeneous materials in nature, and their molecular arrangement has very good uniformity. The arrangement of monatomic structure is essentially the static solution to the multi-bodies problem, which means that there is a close relationship between the static solution to the multi-bodies problem and the uniformity of sequence. These practical examples inspire the author to put forward the assumption that the static solution to multi-bodies problem is a set of low-discrepancy sequence. If *n* particles with the same mass are given in a finite space and the form of force between all pair particles are the same, the static solution to multi-bodies problem will be obtained when they reach the static equilibrium state. If this model is extended to the high-dimensional case, can it produce a low-discrepancy sequence in the high-dimensional space?

Discrepancy is a very effective index to measure the uniformity of sequence. According to Koksma–Hlawka (K-H) inequality, the upper bound of the QMCM error is determined by the discrepancy: the smaller the discrepancy, the smaller the error of the QMCM. However, many scholars [18,26-28] used a minimum distance to measure the uniformity of sequence. This index refers to that the larger the minimum distance of the sequence, the better the uniformity of the sequence. There is no strict mathematical basis for this index, but interestingly, a considerable number of experimental results supported it [18,26-28]. Therefore, the maximized minimum distance is also commonly used in the construction of low-discrepancy sequence.

In fact, maximizing the minimum distance is a common concept in nature and multi-bodies problems. Imagine that there are $n$ predators live in a limited range. The predation ability of these predators is the same, and the distance between the $i$-th and $j$-th predators is denoted by $d_{ij}$, which can be used to measure the size of the predation range of each predator. The predators compete fairly hard to extend their predation range, and meanwhile their predation range are limited by other predators. After a period of competition, it finally runs to a stable state. As each predator has the same predation ability, the size of the predation range of each predator will be the same, and the distances between any pair adjacent predators are almost the same, which corresponds to the maximized minimum distance. This biological model, of course, cannot strictly explain the relationship between the maximized minimum distance and the discrepancy. However, many studies have shown that the larger the minimum distance of the sequence, the smaller the discrepancy of the sequence [26-28]. The author tried to find the corresponding theoretical support of this phenomenon in the relevant knowledge of multi-bodies problem. Our work in Section 2 will construct the equivalence between the maximized minimum distance and the minimum potential energy principle. The examples proposed in Section 5 also show that there is a good positive correlation between the low-discrepancy and the maximized minimum distance or the potential energy. If it is accepted that there is a good positive correlation between the maximized minimum distance and the low-discrepancy, then the low-discrepancy sequence can be constructed by using the maximized minimum distance. Finally, the problem becomes how to construct the low-discrepancy sequence by using the principle of maximized minimum distance, or equivalently, the principle of minimum potential energy, which will be discussed in detail in Sections 3-4.

The arrangement of the monatomic structure is essentially a static solution to multi-bodies problem, which means that there is a close relationship between the static solution to multi-bodies problem and the uniformity of sequence. This is what this paper is interested in. In the past, many studies used the MCM to analyze the evolution of multi-bodies [4,29]. The research in this paper is an inverse way, trying to use the static solution to multi-bodies problem to help to find the sequence with low-discrepancy and improve the accuracy of the MCM, which will establish a channel from the multi-bodies problem to the MCM.

## 2. Relationship between maximized minimum distance and minimum potential energy principle

Consider a sequence $P_{sn} = \{x_1, \cdots, x_n\}$, where $n$ is the number of sequences, $x_i = (x_{i1}, \cdots, x_{is}) \in \Omega := [0,1]^s$, and $s$ represents the spatial dimensionality. The minimum distance of this sequence $P_{s,n}$ is:

$$d_{\min}(P_{s,n}) := \min_{1 \leq i < j \leq n} d_{ij}, \quad d_{ij} = \|x_i - x_j\|. \tag{1}$$

Maximizing the minimum distance means that for a given $n$, it is necessary to find a sequence to meet the following requirement

$$\max_{\forall P_{s,n}} \{d_{\min}(P_{s,n})\} := \max_{\forall P_{s,n}} \left\{ \min_{1 \leq i < j \leq n} \|x_i - x_j\| \right\}. \tag{2}$$

It is not easy to find a sequence with the maximized minimum distance, as the objective function $d_{\min}(P_{s,n})$ is non-smooth. The optimization algorithms based on gradient are not available, and one has to use the method based on the ergodic search or the heuristic intelligent optimization algorithms. Here, we establish a formulation equivalent to Eq. (2), i.e.,

$$\min_{P_n} \left\{ U_\infty : U_\infty := \max_{1 \leq i < j \leq n} \frac{1}{d_{ij}}, \quad x_i, x_j \in P_{s,n} \right\}, \tag{3}$$

and $d_{\min}(P_{s,n}) = U_\infty^{-1}$. To maximize $d_{\min}(P_{s,n})$ is equivalent to minimize $U_\infty$. For a sequence $P_{s,n}$, there is a vector

$$d_* = \left( \frac{1}{d_{12}}, \frac{1}{d_{13}}, \cdots, \frac{1}{d_{1n}}, \frac{1}{d_{23}}, \cdots, \frac{1}{d_{2n}}, \cdots \right), \tag{4}$$

and $U_\infty = \|d_*\|_\infty$. According to the equivalence theory of vector norm [30], we can also use any $p$-norm, i.e.,

$$U_p = \left( \sum_{1 \leq i < j \leq n} \frac{1}{d_{ij}^p} \right)^{\frac{1}{p}}. \tag{5}$$

If $p \to \infty$

$$\lim_{p \to \infty} U_p = U_\infty = \max_{1 \leq i < j \leq n} \frac{1}{d_{ij}}, \tag{6}$$

else if $p = 1$,

$$U_1 = \sum_{1 \leq i < j \leq n} \frac{1}{d_{ij}}. \tag{7}$$

In fact, if each point in the sequence is regarded as a star with mass of 1, the distance between any two stars is $d_{ij}$. Any two stars are affected by the universal gravitation with a gravitational constant of 1, Eq. (7) is the potential energy of the whole galaxy. According to the principle of minimum potential energy in mechanics, the distribution of galaxies satisfies $\min U_1$. When the potential energy of galaxies is the smallest, the distribution of the interaction forces between stars is the most balanced. Therefore, the potential energy can be regarded as an index to measure to the discrepancy of the sequence. Based on the equivalence principle of norm, for any norm $p > 0$, $U_p$ always denotes a potential energy and $U_\infty$ is just a special case. In other words, when the potential energy of the galaxy is the smallest, the stars are most evenly distributed. This is a viewpoint in the sense of physics: for a galaxy in which all the stars have the same mass and are subject to the same form of gravity, if it is static and the resultant force acting on each star is zero, the galaxy distribution should also be uniform. Therefore, potential energy is indeed a measure of the uniformity of galaxy distribution, and maximizing minimum distance is equivalent to minimizing potential energy $U_\infty$. The question now is how to get this evenly distributed galaxy?

Before answering this question, we first give a brief introduction to the definition of the distance. In the Euclidean space, the distance is usually defined as:

$$d_{\mathrm{E},ij} = \|\boldsymbol{x}_i - \boldsymbol{x}_j\|_2 := \sqrt{\sum_{k=1}^{s}(x_{ik} - x_{jk})^2} . \tag{8}$$

In a cube $\Omega := [0,1]^s$, we can also adopt the circular distance [26-28]

$$d_{\mathrm{T},ij} = \|\boldsymbol{x}_i - \boldsymbol{x}_j\|_{\mathrm{T}} := \sqrt{\sum_{i=1}^{s}\left(\min\left\{|x_{ik} - x_{jk}|, 1 - |x_{ik} - x_{jk}|\right\}\right)^2} . \tag{9}$$

Using the equivalence theory of vector norm, we can further extend the above distance. Obviously

$$\min\left\{|x_{ik} - x_{jk}|, 1 - |x_{ik} - x_{jk}|\right\} = \frac{1}{\max\left\{\dfrac{1}{|x_{ik} - x_{jk}|}, \dfrac{1}{1 - |x_{ik} - x_{jk}|}\right\}} . \tag{10}$$

Thus, according to the equivalence principle of norm, a new distance can be defined:

$$d_{q,ij} := \sqrt{\sum_{i=1}^{s}\left(\frac{1}{\dfrac{1}{|x_{ik} - x_{jk}|^q} + \dfrac{1}{(1 - |x_{ik} - x_{jk}|)^q}}\right)^{\frac{2}{q}}} = \sqrt{\sum_{k=1}^{s}\frac{|x_{ik} - x_{jk}|^2 (1 - |x_{ik} - x_{jk}|)^2}{\left[(1 - |x_{ik} - x_{jk}|)^q + |x_{ik} - x_{jk}|^q\right]^{\frac{2}{q}}}} . \tag{11}$$

Obviously, Eq. (11) is an extension of Eq. (9). When $q = \infty$, Eq. (11) degenerates to Eq. (9). For convenience of marking, we denote the potential energy as $U_{q,p}$. When $q > 0$ and $q$ is a real number, it indicates that the distance adopts Eq. (11), and when $q \equiv \mathrm{'E'}$, it indicates that the distance adopts Eq. (8).

# 3. Dynamical evolutionary model

This section will establish a dynamical evolutionary model (DEM) to find out the sequence $P_{s,n}$ with the minimum potential energy. Assuming that a sequence $P_{s,n} = \{x_1, \cdots, x_n\}$ is distributed in the cube $\Omega = [0,1]^s$, each point is regarded as a star with mass of $m$, and there is interaction force between any pair points to form the potential energy $U_{q,p}$. According to the analytical mechanics, the action and Lagrange function of the whole sequence are as follows:

$$S = \int_0^t L \mathrm{d}\tau, \quad L = \frac{1}{2} m \sum_{i=1}^{n} \sum_{k=1}^{s} \dot{x}_{ik}^2 - G \left( \sum_{1 \leq i < j \leq n} \frac{1}{d_{q,ij}^p} \right)^{\frac{1}{p}}, \tag{12}$$

where $G$ is the gravitational constant. According to Hamilton variational principle [31],

$$\delta S = \int_0^t \left[ -\sum_{i=1}^{n} \sum_{k=1}^{s} \delta x_{ik} m \ddot{x}_{ik} + G \left( \sum_{1 \leq i < j \leq n} \frac{1}{d_{q,ij}^p} \right)^{\frac{1}{p}-1} \sum_{1 \leq i < j \leq n} \frac{\delta d_{q,ij}}{d_{q,ij}^{p+1}} \right] \mathrm{d}\tau = 0, \tag{13}$$

where

$$\delta d_{q,ij} = \frac{1}{d_{q,ij}} \sum_{k=1}^{s} a_{ijk} \left( \delta x_{ik} - \delta x_{jk} \right). \tag{14}$$

If $q = \text{'E'}$,

$$a_{ijk} = x_{ik} - x_{jk}; \tag{15}$$

and if $q > 0$,

$$a_{ijk} = \frac{\mathrm{sgn}(x_{ik} - x_{jk}) \Delta_{ijk} (1 - \Delta_{ijk}) \left[ (1 - \Delta_{ijk})^{q+1} - \Delta_{ijk}^{q+1} \right]}{\left[ (1 - \Delta_{ijk})^q + \Delta_{ijk}^q \right]^{1+\frac{2}{q}}}, \quad \Delta_{ijk} = |x_{ik} - x_{jk}|. \tag{16}$$

Substituting Eq. (14) into Eq. (13),

$$\int_0^t \left[ -\sum_{k=1}^{s} \sum_{i=1}^{n} \delta x_{ik} m \ddot{x}_{ik} + G \left( \sum_{1 \leq i < j \leq n} \frac{1}{d_{q,ij}^p} \right)^{\frac{1}{p}-1} \sum_{k=1}^{s} \sum_{i=1}^{n-1} \sum_{\substack{j=1 \\ j \neq i}}^{n} \frac{a_{ijk} \delta x_{ik}}{d_{q,ij}^{p+2}} \right] \mathrm{d}\tau = 0. \tag{17}$$

According to the above equation, and considering the arbitrariness of $\delta x_{ik}$, it can be obtained that

$$m \ddot{x}_{ik} + f_{ik} = 0, \quad f_{ik} = -G \left( \sum_{1 \leq i < j \leq n} \frac{1}{d_{q,ij}^p} \right)^{\frac{1}{p}-1} \sum_{\substack{j=1 \\ j \neq i}}^{n} \frac{a_{ijk}}{d_{q,ij}^{p+2}}, \quad 1 \leq i \leq n, \; 1 \leq k \leq s, \tag{18}$$

which is a non-linear differential equation describing the dynamic evolution of all the $n$ stars under the action of potential energy $U_{q,p}$. If $p = \infty$,

$$f_{ik} = \begin{cases} -G \dfrac{a_{ijk}}{d_{\min}^3}, & \text{if } d_{p,ij} = d_{\min} := \min_{1 \le i < j \le n} d_{p,ij} \\ 0, & \text{otherwise} \end{cases}. \qquad (19)$$

In order to obtain the static solution, we apply an artificial damping force to each star, so the Eq. (18) is changed to

$$m\ddot{x}_{ik} + c\dot{x}_{ik} + f_{ik} = 0, \quad 1 \le i \le n, \quad 1 \le k \le s. \qquad (20)$$

To solve the above equation, the initial conditions need to be given. Assuming that a sequence $P_{s,n}^{(0)} = \{x_1(0), \cdots, x_n(0)\}$ is given at the initial time and the initial velocity is zero, then the evolution calculation is carried out according to Eq. (20). Due to the effect of damping, the velocity $\dot{x}_{ik}$ and the acceleration $\ddot{x}_{ik}$ tend to zero with time $t \to \infty$. Finally the static sequence distribution is obtained, and the potential energy of the static distribution is the smallest.

## 4. Evolutionary algorithm

Although Eq. (20) is a non-linear differential equation, there are many mature algorithms for it [32]. Here, we use the central difference method to solve it, and take the time step as $\Delta t$. It is assumed that the initial conditions are known as:

$$x_{ik}(0) = x_{ik}^{(0)}, \quad \dot{x}_{ik}(0) = 0, \qquad (21)$$

and the initial acceleration is

$$\ddot{x}_{ik}(0) = m^{-1} f_{ik}\big|_{t=0}. \qquad (22)$$

The solution at $\Delta t$ can be approximated by using the Taylor series:

$$x_{ik}(\Delta t) = x_{ik}^{(1)} = x_{ik}^{(0)} + \frac{1}{2}\ddot{x}_{ik}^{(0)}\Delta t^2 + O(\Delta t^3). \qquad (23)$$

With $x_{ik}^{(0)}$ and $x_{ik}^{(1)}$, Eq. (20) can be solved numerically by using the central difference method [32], i.e.,

$$m\frac{x_{ik}^{(j+1)} + x_{ik}^{(j-1)} - 2x_{ik}^{(j)}}{\Delta t^2} + c\frac{x_{ik}^{(j+1)} - x_{ik}^{(j-1)}}{2\Delta t} + f_{ik}^{(j)} = 0, \qquad (24)$$

where the superscript $(j)$ represents the value at time $t = j \times \Delta t$. Rearranging Eq. (24) gives

$$x_{ik}^{(j+1)} = \frac{A_1 x_{ik}^{(j)} - A_2 x_{ik}^{(j-1)} - f_{ik}^{(j)}}{A_0}, \qquad (25)$$

where

$$A_0 = \frac{m}{\Delta t^2} + \frac{c}{2\Delta t}, \quad A_1 = \frac{2m}{\Delta t^2}, \quad A_2 = \frac{m}{\Delta t^2} - \frac{c}{2\Delta t}. \qquad (26)$$

According to Eq. (25), the sequence of at each time step can be obtained, and the sequence will approach to the final static solution as the iteration progresses. Although the central difference method is a common calculation format, there are still some problems to be further discussed when it is used to the proposed multi-bodies dynamical model.

## 4.1 Selection of evolutionary parameters

In the above evolutionary algorithm, the parameters $m$, $c$, $G$, and $\Delta t$ need to be given. In the actual calculation, the selection of these parameters will influence the convergence speed of the iteration. This section discusses how to select these iteration parameters. To facilitate analysis, we rewrite Eq. (20) in the following vector form

$$m\ddot{X} + c\dot{X} + f(X) = 0, \quad f(X) = \frac{\partial U_{q,p}(X)}{\partial X}, \tag{27}$$

where $X = (x_{11}, \cdots, x_{1s}, \cdots, x_{n1}, \cdots, x_{ns})^{\mathrm{T}}$. Using the central difference method, the above equation can be approximated as

$$X_{j+1} = \left(\frac{m}{\Delta t^2} + \frac{c}{2\Delta t}\right)^{-1} \frac{2m}{\Delta t^2} X_j - \left(\frac{m}{\Delta t^2} + \frac{c}{2\Delta t}\right)^{-1} \left(\frac{m}{\Delta t^2} - \frac{c}{2\Delta t}\right) X_{j-1} - \left(\frac{m}{\Delta t^2} + \frac{c}{2\Delta t}\right)^{-1} f(X_j). \tag{28}$$

It can be seen from Eq. (28) that if $X_j$ tends to a fixed value $X_*$ when $j \to \infty$, there must be

$$X_* = \left(\frac{m}{\Delta t^2} + \frac{c}{2\Delta t}\right)^{-1} \frac{2m}{\Delta t^2} X_* - \left(\frac{m}{\Delta t^2} + \frac{c}{2\Delta t}\right)^{-1} \left(\frac{m}{\Delta t^2} - \frac{c}{2\Delta t}\right) X_* - \left(\frac{m}{\Delta t^2} + \frac{c}{2\Delta t}\right)^{-1} f(X_*). \tag{29}$$

It can be easily seen from the above equation that $f(X_*) = \dfrac{\partial U_{q,p}(X_*)}{\partial X_*} = 0$, which implies if the iterative format (28) converges, $U_{q,p}(X_*)$ must be the minimum potential energy. Then how to select parameters to ensure the convergence? Let $u_{j+1} = \begin{pmatrix} X_j^{\mathrm{T}} & X_{j+1}^{\mathrm{T}} \end{pmatrix}^{\mathrm{T}}$ be regarded as a state vector, Eq. (28) and $X_j = X_j$ jointly give a mapping relationship $u_{j+1} = \varphi(u_j)$ from $u_j$ to $u_{j+1}$, and the Jacobi matrix of the mapping is

$$J(u_j) = \begin{bmatrix} \mathbf{0} & I \\ -\left(\dfrac{m}{\Delta t^2} + \dfrac{c}{2\Delta t}\right)^{-1}\left(\dfrac{m}{\Delta t^2} - \dfrac{c}{2\Delta t}\right)I & \left(\dfrac{m}{\Delta t^2} + \dfrac{c}{2\Delta t}\right)^{-1}\left(\dfrac{2m}{\Delta t^2}I - K(u_j)\right) \end{bmatrix}, \tag{30}$$

where

$$K(u_j) = \frac{\partial f}{\partial X_j^{\mathrm{T}}} = \frac{\partial^2 U(X_j)}{\partial X_j \partial X_j^{\mathrm{T}}}. \tag{31}$$

According to the fixed point iteration theorem [30], the convergence of the iterative method mainly depends on the eigenvalues of the Jacobi matrix at the solution $X_*$. If the moduluses of all the eigenvalues are not greater than 1, the iteration is stable and will not diverge. The eigen problem of Jacobi matrix can be written as:

$$J(u_*)\begin{pmatrix}y_1\\y_2\end{pmatrix}=\lambda\begin{pmatrix}y_1\\y_2\end{pmatrix},\quad u_*=\begin{pmatrix}X_*^{\mathrm{T}} & X_*^{\mathrm{T}}\end{pmatrix}^{\mathrm{T}}. \tag{32}$$

Substituting Eq. (30) into Eq. (32) yields:

$$K(u_*)y_1=\mu y_1,\quad \mu=\frac{2m}{\Delta t^2}-\lambda\left(\frac{m}{\Delta t^2}+\frac{c}{2\Delta t}\right)-\frac{1}{\lambda}\left(\frac{m}{\Delta t^2}-\frac{c}{2\Delta t}\right), \tag{33}$$

where $\mu$ is the eigenvalue of $K(u_*)$. $U_{q,p}$ is a convex function near the minimum $X_*$ and $K(u_*)$ is a symmetric positive definite matrix, which means $\mu>0$. According to Eq. (32),

$$\lambda=\frac{-\left(\mu-\frac{2m}{\Delta t^2}\right)\pm\sqrt{\mu^2-\mu\frac{4m}{\Delta t^2}+\frac{c^2}{\Delta t^2}}}{2\left(\frac{m}{\Delta t^2}+\frac{c}{2\Delta t}\right)}. \tag{34}$$

In order to ensure the iterative convergence, we must ensure $|\lambda|<1$. For simplicity, let

$$x=\mu,\quad y=\frac{2m}{\Delta t^2},\quad z=\frac{c}{\Delta t},\quad x,y,z\geq 0, \tag{35}$$

Eq. (34) can be rewritten as

$$\lambda=\frac{-(x-y)\pm\sqrt{x^2-2xy+z^2}}{y+z}. \tag{36}$$

Now the question becomes how to choose $y$ and $z$ to ensure $|\lambda|<1$. For the convenience of discussion, we denote the minimum and maximum eigenvalues of $K(u_*)$ as $\mu_0$ and $\mu_1$ respectively. Obviously, different $y$ and $z$ will lead to different $\lambda$ and need to be discussed in different situations. For example, when $y<z$, $x^2-2xy+z^2>0$; when $y>z$, $x^2-2xy+z^2$ may be greater than or less than zero. After a tedious calculation based on the basic knowledge of advanced mathematics, we can get the following relationship:

if $y\geq z>0$

$$\max|\lambda|=\begin{cases}\dfrac{y-x+\sqrt{x^2-2xy+z^2}}{y+z}, & x<y-\sqrt{y^2-z^2}\\[2mm]\dfrac{y-x}{y+z}, & y-\sqrt{y^2-z^2}<x\leq y\\[2mm]\dfrac{x-y}{y+z}, & y<x<y+\sqrt{y^2-z^2}\\[2mm]\dfrac{x-y+\sqrt{x^2-2xy+z^2}}{y+z}, & y+\sqrt{y^2-z^2}<x\end{cases}; \tag{37}$$

else if $0<y<z$

$$\max|\lambda| = \begin{cases} \dfrac{y - x + \sqrt{x^2 - 2xy + z^2}}{y + z}, & x < y \\ \dfrac{x - y + \sqrt{x^2 - 2xy + z^2}}{y + z}, & y \le x \end{cases}. \tag{38}$$

According to the stability condition, $\max|\lambda| < 1$. Combining this condition with Eqs. (37)-(38), the value ranges of $y$ and $z$ satisfying $|\lambda| < 1$, which are also called stability region, are obtained and drawn in Fig. 1. The abscissa is $z$ and the ordinate is $y$. The colored regions represents the stability regions in which $|\lambda| < 1$. It is obvious that the condition satisfying $|\lambda| < 1$ is $y > \dfrac{\mu_1}{2}$. We have studied the effect of $y$ and $z$ on $\max|\lambda|$. There are three lines in Fig. 1: line $l_1$ is $y = z$, line $l_2$ is $y = \dfrac{z^2}{2\mu_0} + \dfrac{\mu_0}{2}$, and line $l_3$ is $y = \dfrac{z^2}{2\mu_1} + \dfrac{\mu_1}{2}$. Lines $l_2$, $l_3$ and $y = \dfrac{\mu_0 + \mu_1}{2}$ jointly divided the stability region into four sub-regions (a)-(d), as shown in Fig. 1, in which the arrow drawn in Fig. 1 indicates the attenuation path of $\max|\lambda|$. In stability region (a), $\max|\lambda|$ decreases along $z$ and along $-y$; in stability region (b), $\max|\lambda|$ decreases along $-y$ and along $-z$; in stability region (c), $\max|\lambda|$ decreases along $y$ and along $z$; in stability region (d), $\max|\lambda|$ decreases along $y$ and along $-z$. When $z = \sqrt{\mu_0\mu_1}$ and $y = \dfrac{\mu_0 + \mu_1}{2}$, $\max|\lambda| = \dfrac{\sqrt{\mu_1} - \sqrt{\mu_0}}{\sqrt{\mu_1} + \sqrt{\mu_0}}$ reaches the global minimum. The global minimum point is marked by the red dot in Fig. 1. When $z = y = \dfrac{\mu_0 + \mu_1}{2}$, $\max|\lambda| = \dfrac{\mu_1 - \mu_0}{\mu_1 + \mu_0}$ is a local minimum, this point is marked as a blue point in Fig. 1. The two intersections of lines $l_1$ and $l_2$ and of lines $l_1$ and $l_3$ are the other two local minimum points.

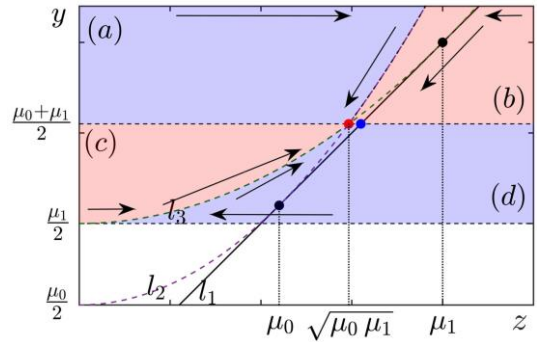

Figure 1 Schematic diagram of stability region in which $|\lambda| < 1$

To obtain the fastest convergence speed, we must take $z = \sqrt{\mu_0\mu_1}$ and $y = \dfrac{\mu_0 + \mu_1}{2}$. However the exact values of $\mu_0$ and $\mu_1$ are unknown in the actual calculation, and hence the fastest convergence speed cannot obtained. Considering $\mu_0$ and $\mu_1$ are the eigenvalues of $\boldsymbol{K}(\boldsymbol{u}_*) = \dfrac{\partial^2 U(\boldsymbol{X}_*)}{\partial \boldsymbol{X}_* \partial \boldsymbol{X}_*^{\mathrm{T}}}$, in actual calculation, we take

$$\mu_0 \approx \frac{U(\boldsymbol{X}_0)}{\boldsymbol{X}_0^{\mathrm{T}} \boldsymbol{X}_0}, \quad \mu_1 \approx \kappa \mu_0, \tag{39}$$

where $\boldsymbol{X}_0$ is the initial sequence and $\kappa$ is a given constant. According to the above equation, we have

$$z = \frac{U(\boldsymbol{X}_0)}{\boldsymbol{X}_0^{\mathrm{T}} \boldsymbol{X}_0} \sqrt{\kappa}, \quad y = \frac{1+\kappa}{2} \frac{U(\boldsymbol{X}_0)}{\boldsymbol{X}_0^{\mathrm{T}} \boldsymbol{X}_0}. \tag{40}$$

Combining Eq. (35) with Eq. (40) yields

$$m = \frac{1+\kappa}{4} \frac{U(\boldsymbol{X}_0)}{\boldsymbol{X}_0^{\mathrm{T}} \boldsymbol{X}_0} \Delta t^2, \quad c = \frac{U(\boldsymbol{X}_0)}{\boldsymbol{X}_0^{\mathrm{T}} \boldsymbol{X}_0} \sqrt{\kappa} \Delta t. \tag{41}$$

In order to ensure the convergence of the iteration, $y > 0.5\mu_1$ needs to be satisfied, but $\mu_1$ cannot be given exactly. Therefore, $\kappa$ in Eq. (40) cannot be accurately defined. In this paper, a trial calculation method is adopted: give a $\kappa$ (such as $\kappa = 100$), and then perform the iterations some steps; if the potential energy $U(\boldsymbol{X}_j)$ keep decreasing with the iteration, the given $\kappa$ should be adopted, otherwise $\kappa$ should be enlarged. It can be seen from the above analysis that $\Delta t$ and $G$ do not really affect the convergence of the iteration. In this report, $\Delta t$ and $G$ are set to be 0.1 and 1, respectively.

## 4.2 Points walking out of the bound

During the evolution, all the points walk under the action of the gravitation, damping, and inertia forces. Although all the points are is required to fall within the cube $\Omega := [0,1]^s$, there may be a point (e.g., the $i$-th point) with the position $\boldsymbol{x}_i^{(j)}$ walk out of $\Omega$ at certain time $t_j$. At this time, $x_{ik}^{(j)} > 1$ or $x_{ik}^{(j)} < 0$ appears in a coordinate of $\boldsymbol{x}_i^{(j)}$. If the point is out of bounds, it needs to be corrected manually, otherwise it will lead to the divergence of the iteration. In this paper, the periodic condition is used to deal with the points out of the bound. The two-dimensional space shown in Fig. 2 is taken as an example.

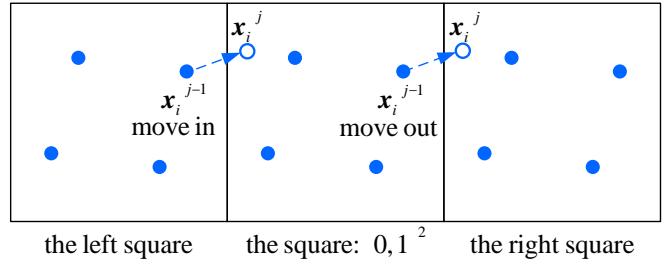

Figure 2 Schematic diagram of periodic arrangement of two-dimensional squares

Figure 2 shows several points distributed in a two-dimensional square $\Omega := [0,1]^2$. Suppose that three squares are arranged side by side, the distribution of sequence in each square is the same, and the evolution of sequence in each square is also exactly the same. Let the $i$-th point in the middle square keeps moving to the right. If this point moves out of the middle square at time $t_j$ and enters the right square, there must be a point at the same position in the left square which will move into this middle square at the same time $t_j$. Therefore, the moving-in point can be used to replace the moving-out point for the next evolutionary iterative step. This idea can be performed by changing the coordinate $x_{ik}$ to be

$$x_{ik} = \begin{cases} x_{ik} - [x_{ik}], & \text{if } x_{ik} > 1 \\ x_{ik} - [x_{ik}] + 1, & \text{if } x_{ik} < 0 \\ x_{ik}, & \text{otherwise} \end{cases}, \tag{42}$$

where $[x_{ik}]$ represents the integer closest to $x_{ik}$.

## 4.3 Unbiased sequence

If a set of random variables $x \in \Omega := [0,1]^s$ obeys the uniform distribution, the mean of $x$ must be 0.5. If the sample mean of a sequence satisfies

$$\frac{1}{n}\sum_{i=1}^{s} x_{ik} = 0.5, \tag{43}$$

then the sample sequence is said to be unbiased. The existing low-discrepancy sequences generated based on the number theory are often not unbiased. In this sub-section, we propose a restarting technique to make the sequence generated based on the proposed DEM to be unbiased. The restarting technique involves a concept of absolute mean difference $\Delta P$ defined by

$$\Delta P = \left\| \frac{1}{n}\sum_{i=1}^{n} x_i - 0.5 \right\|_\infty, \tag{44}$$

which is actually the difference between the sample means of the sequence and the exact mean. At each iterative step, if the sequence is obtained, the absolute difference $\Delta P$ will also be computed.

If $\Delta P$ is less than a given threshold $\varepsilon_p$, the evolutionary calculation continues; if $\Delta P$ is larger than $\varepsilon_p$, the evolutionary iteration is ended and the sequence is adjusted as follows:

$$x_i := x_i - \frac{1}{n}\sum_{i=1}^{n} x_i + 0.5. \tag{45}$$

Taking the adjusted sequence as the initial point, the evolutionary iteration is restarted again.

## 5. Numerical experiment

## 5.1 Relationship between potential energy and discrepancy in evolution

As mentioned in Section 2, the potential energy can be regarded as an index to measure the uniformity of the sequence. Obviously, this viewpoint is obtained from a physical point of view, and is actually an assumption without strict mathematical foundation. If this viewpoint is correct, the potential energy will decreases gradually with the increase of iterative steps, and the discrepancy of the sequence will decreases synchronously. In this subsection, we will design some experiments to verify this viewpoint. At present, there are many indexes to measure the discrepancy of sequence. Here two commonly used indexes, namely the star L2-discrepancy and the modified L2-discrepancy, which are defined as

$$D_{L_2}(P_{s,n})^2 = \left(\frac{1}{3}\right)^s - \frac{2}{n}\sum_{i=1}^{n}\prod_{j=1}^{s}\left(\frac{1-x_{ij}^2}{2}\right) + \frac{1}{n^2}\sum_{i,l=1}^{n}\prod_{j=1}^{s}\left[1-\max\left(x_{ij},x_{lj}\right)\right] \tag{46}$$

and

$$D_{MD}(P_{s,n})^2 = \left(\frac{4}{3}\right)^s - \frac{2^{1-s}}{n}\sum_{i=1}^{n}\prod_{j=1}^{s}\left(3-x_{ij}^2\right) + \frac{1}{n^2}\sum_{i,l=1}^{n}\prod_{j=1}^{s}\left[2-\max\left(x_{ij},x_{lj}\right)\right], \tag{47}$$

are adopted.

**5.1.1 Sequences with $s=2$**

Firstly, we study the case of $s=2$, and take different sample numbers $n=10, 20, 40, 80, 160$ and 320 to generate different sequences $P_{2,n}$. The potential energy is calculated by $U_{1,1}$ and $U_{E,\infty}$ respectively. $U_{1,1}$ and $U_{E,\infty}$ adopt the distances shown in Eq. (11) and Eq. (8), respectively. The evolutionary algorithm developed in Section 4 is used for calculation. For each sequence $P_{2,n}$, the random sampling method is used to generate the initial sequence $P_{2,n}^{(0)}$. The total number of the evolutionary iterative steps is set to be 1000, and hence 1000 sequences $P_{2,n}^{(j)}$ can be obtained, where $j=1,\cdots,1000$. For each sequence $P_{2,n}^{(j)}$, we have four indexes, i.e., two potential energies $U_{1,1}\left(P_{2,n}^{(j)}\right)$ and $U_{E,\infty}\left(P_{2,n}^{(j)}\right)$, and two discrepancies $D_{L_2}\left(P_{2,n}^{(j)}\right)$ and $D_{MD}\left(P_{2,n}^{(j)}\right)$. The Spearman correlation coefficients between the potential energies and the discrepancies are calculated to show that the potential energy is truly an index to measure the uniformity of the sequence from a statistical perspective. For the convenience of discussion, the correlation coefficients between $U_{1,1}\left(P_{2,n}^{(j)}\right)$ and $D_{L_2}\left(P_{2,n}^{(j)}\right)$, $U_{1,1}\left(P_{2,n}^{(j)}\right)$ and $D_{L_2}\left(P_{2,n}^{(j)}\right)$ are recorded as $\rho_{1,L_2}$ and $\rho_{1,MD}$ respectively; the correlation coefficients between $U_{E,\infty}\left(P_{2,n}^{(j)}\right)$ and $D_{L_2}\left(P_{2,n}^{(j)}\right)$, $U_{E,\infty}\left(P_{2,n}^{(j)}\right)$ and $D_{L_2}\left(P_{2,n}^{(j)}\right)$ are recorded as $\rho_{E,L_2}$ and $\rho_{E,MD}$ respectively. The Spearman correlation coefficients for all sequences are listed in Table 1.

Table 1 Spearman correlation coefficients $\rho_{1,L_2}$, $\rho_{1,MD}$, $\rho_{E,L_2}$ and $\rho_{E,MD}$ of two-dimensional sequences with different numbers. Except that the $\rho_{1,MD}$ corresponding $p$-value of $n=20$ is 0.1631, the $p$-value of other cases is less than 1e-15.

| $n$ | 10 | 20 | 40 | 80 | 160 | 320 |
| --- | --- | --- | --- | --- | --- | --- |
| $\rho_{1,L_2}$ | 0.9970 | -0.6550 | -0.8378 | 0.9342 | 0.5750 | 0.9625 |
| $\rho_{1,MD}$ | 0.9995 | -0.0441 | -0.8290 | 0.9544 | 0.5700 | 0.9495 |
| $\rho_{E,L_2}$ | 0.8930 | 0.9485 | 0.7048 | 0.6885 | 0.7953 | 0.7639 |
| $\rho_{E,MD}$ | 0.8917 | 0.9460 | 0.6971 | 0.6968 | 0.9733 | 0.9547 |

It is obvious from the results in Table 1 that there are significant positive correlations between the potential energies and the star L2-discrepancies, or the modified L2-discrepancies in the evolution processes of most sequences, which indicates that the potential energy and the discrepancy have the same change trend in the evolutionary process. For the two sequences $P_{2,20}$ and $P_{2,40}$, the Spearman correlation coefficients between $U_{1,1}$ and $D_{L_2}$ or $D_{MD}$ are negative, which is mainly because the iteration is restarted in the evolutionary process, which results in the fluctuation of the discrepancy, as shown in subsection 4.3,. In fact, the change trend of the discrepancy is always consistent with that of the potential energy for these two cases. Taking $P_{2,20}$ as an example, the evolution curves of $D_{L_2}$, $D_{MD}$, and $U_{1,1}$ are shown in Fig. 3. It can be seen that in the initial stage of the evolution, the discrepancy fluctuates due to the restart of the iteration, but the overall change trend of $D_{L_2}$ or $D_{MD}$ is very close to that of $U_{1,1}$. Taken overall, the change trend of the discrepancy is for all the considered sequences consistent with that of the potential energy.

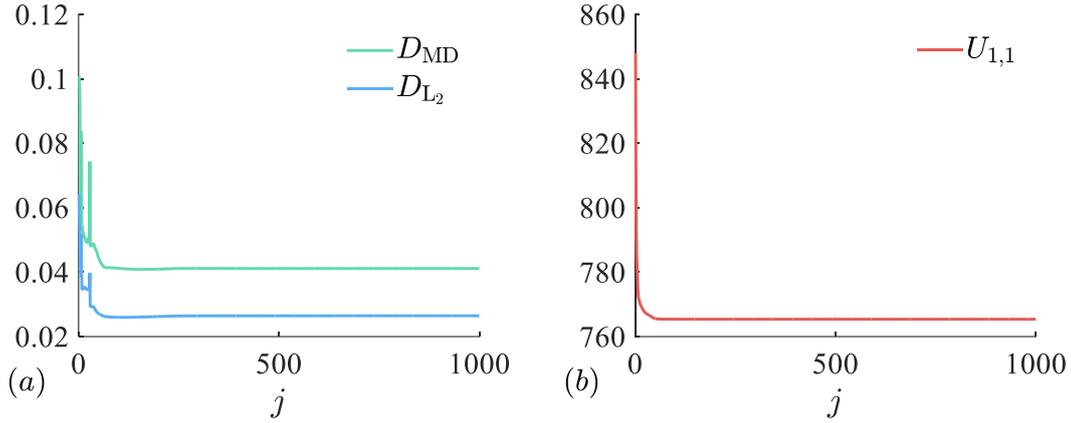

Figure 3 Discrepancy and potential energy in the evolution of $P_{2,20}$: (a) $D_{L_2}\left(P_{2,20}^{(j)}\right)$ and $D_{MD}\left(P_{2,20}^{(j)}\right)$; (b) $U_{1,1}\left(P_{2,20}^{(j)}\right)$

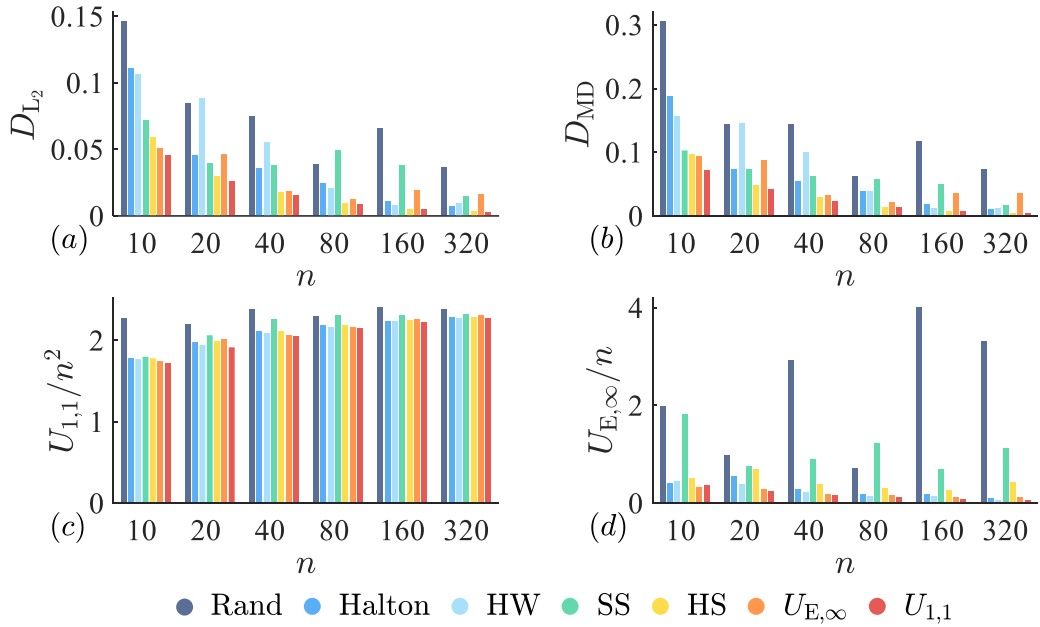

Figure 4 Comparison of the uniformity indexes of the sequences generated by different methods: (a) $D_{L_2}$; (b) $D_{MD}$; (c) $U_{1,1}$; (d) $U_{E,\infty}$

Next the sequences generated by different methods are compared in terms of $D_{L_2}$, $D_{MD}$, $U_{1,1}$ and $U_{E,\infty}$. The sequences we compare include the random sampling sequence generated by using $\text{rand}(2,n)$, the Halton sequence, the Halton sequence combined with scrambling method (HS), the Sobol sequence combined with scrambling method (SS), the Hua Wang sequence (HW), the sequence generated based on the DEM with $U_{1,1}$, and the sequence generated based on the DEM with $U_{E,\infty}$. We calculated $D_{L_2}$, $D_{MD}$, $U_{1,1}$ and $U_{E,\infty}$ of the sequences generated by different methods, and plotted them in Figs. 4(a)-4(d). It can be seen from Fig. 4 that the DEM with $U_{E,\infty}$ or $U_{1,1}$ can generate a low-discrepancy sequence and its discrepancy is much smaller than that of the random sampling sequence. The uniformity of the sequence generated based on $U_{1,1}$ is better than that generated based on $U_{E,\infty}$. Compared with other sequences, the discrepancy and potential energy of the sequences generated by using the DEM with $U_{1,1}$ are smaller.

According to Figs 3-4 and Table 1, the potential energy can be used as an index to measure the uniformity of sequences, and the static solutions of the multi-points dynamical model are truly low-discrepancy sequences.

### 5.1.2 Sequences with $n=100$

We then set $n=100$ and give different dimensions, i.e., $s=2$, 4, 6, 8, and 10. $U_{1,1}$ and $U_{E,\infty}$ are also used to calculate. The total number of the evolutionary iterative steps is also set to be 1000. The Spearman correlation coefficients and $p$-values between the potential energies and the discrepancies of all the sequences are also calculated and listed in Table 2. Because the $p$-values in all cases are less than 1e-15, they are not listed. The calculation results again show that for almost all sequences with the same number of samples but different dimensions, there is a significant positive correlation between the potential energy and the discrepancy, indicating that the potential energy and the discrepancy have the same change trend in the evolutionary process. For the two sequences $P_{4,100}$ and $P_{6,100}$, the Spearman correlation between $U_{1,1}$ and $D_{L_2}$ is weak, while the Spearman correlation between $U_{1,1}$ and $D_{MD}$ is strong. This is mainly because the restart of the iteration in the evolutionary process, resulting in the fluctuation of $D_{L_2}$. Taken overall, there is an obvious positive correlation between the potential energy and the discrepancy in the evolutionary process of sequences with different dimensions.

Table 2 Spearman correlation coefficients $\rho_{1,L_2}$, $\rho_{1,MD}$, $\rho_{E,L_2}$ and $\rho_{E,MD}$ of two-dimensional sequences with different sample numbers. $p$-value is in all cases less than 1e-15.

| $s$ | 2 | 4 | 6 | 8 | 10 |
|---|---|---|---|---|---|
| $\rho_{1,L_2}$ | 0.8128 | 0.5967 | 0.6048 | 0.9629 | 0.6343 |
| $\rho_{1,MD}$ | 0.8387 | 0.8786 | 0.8814 | 0.9944 | 0.7849 |
| $\rho_{E,L_2}$ | 0.7484 | 0.8488 | 0.9386 | 0.9320 | 0.9625 |
| $\rho_{E,MD}$ | 0.8005 | 0.8679 | 0.9470 | 0.9898 | 0.9855 |

Again, we compare the uniformities of the sequences generated by different methods. The sequences used include the random sampling sequence, the Halton sequence, the HS sequence, the SS sequence, the HW sequence, the sequence generated using the DEM with $U_{1,1}$, and the sequence generated using the DEM with $U_{E,\infty}$. $D_{L_2}$, $D_{MD}$, $U_{1,1}$ and $U_{E,\infty}$ of the sequences generated by different methods are calculated and shown in Figs. 5(a)-5(d). A conclusion similar to that in proposed 5.1.1 can be obtained from Fig. 5: for different dimensions, the DEM with $U_{1,1}$ or $U_{E,\infty}$ can generate low-discrepancy sequences; the uniformity of the sequence generated based on $U_{1,1}$ is better than that generated based on $U_{E,\infty}$, and is also better than that generated by other methods.

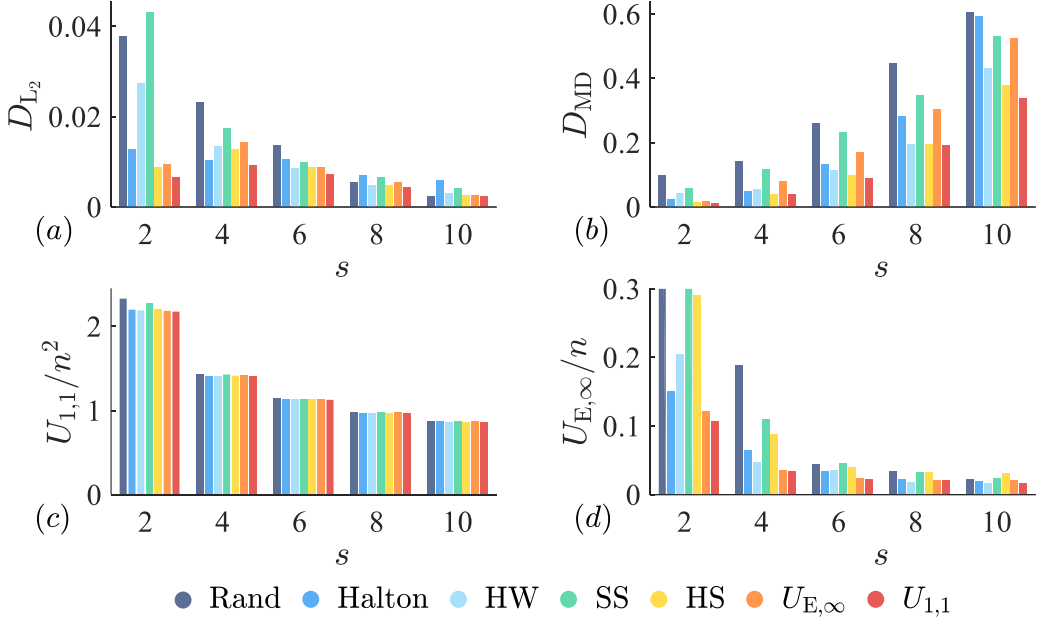

Figure 5 Comparison of the uniformity indexes of the sequences generated by different methods:
(a) $D_{L_2}$; (b) $D_{MD}$; (c) $U_{1,1}$; (d) $U_{E,\infty}$

Figure 5 and Table 2 demonstrate again that the potential energy can be used as an index to measure the uniformity of sequences, and the static solutions of the multi-points dynamical model are truly low-discrepancy sequences.

## 5.2 Generate a series of low discrepancy sequences

Using the proposed DEM with the potential energy $U_{q,p}$, we have generated 127 sequences $P_{s,n}$ with $s \in \{1, 2, \cdots, 20\}$, $n \in \{10, 20, 50, 100, 200, 400, 800, 1500\}$ and $n \geq 2s$. All these sequences can be downloaded from https://rocewea.com/2.html. The evolutionary parameters used to generate these sequences and the comparison of these generated sequences with other methods are presented in the following subsections.

**5.2.1 Selection of $q$ and $p$**

In the practical calculation, we have found that different $q$ and $p$ can generate different the sequences with different potential energies and discrepancies. It is nature to expect that a pair of $q$ and $p$ will be chosen for a given $(n, s)$ to make the discrepancy $D_{MD}(P_{s,n})$ be as small as possible during the process of minimizing the potential energy. In order to achieve this goal, we use the uniform test design [33] and take 10 sequences in $(n, s) \in [s, 1500] \times [0, 20]$ (see rows 2 and 3 of Table 3). For each $(n_i, s_i)$, we will find a pair of $(\hat{p}_i, \hat{q}_i)$ that makes $D_{MD}(P_{s_i,n_i})$ be smallest. Based on $(n_i, s_i)$ and $(\hat{p}_i, \hat{q}_i)$, we will establish the nonlinear regression model of $(q, p)$ and $(n, s)$.

In order to find $(\hat{p}_i, \hat{q}_i)$ corresponding to $(n_i, s_i)$, we use ten different pairs of $(p_k, q_k) \in (0, 1]$ ($k = 1, \cdots, 10$) to generate ten sequences $P_{s_i,n_i}(p_k, q_k)$ by the proposed DEM with $U_{q_k,p_k}$. The $(p_k, q_k)$ is selected from the sequence $P_{2,10}$ generated in subsection 5.1.1, and listed in rows 4 and 5 of Table 3. For each pair of $(p_k, q_k)$, the evolutionary iteration is performed with 1000 iterative steps, and the discrepancy $D_{MD}(p_k, q_k)$ of the sequence at the last step is calculated. Finally, we choose $(p_k, q_k)$ corresponding to the minimum $\min_{1 \leq k \leq 10}\{D_{MD}(p_k, q_k)\}$ as the $(\hat{p}_i, \hat{q}_i)$ of $(n_i, s_i)$. After calculation, the results are listed in rows 6 and 7 of Table 3.

Table 3 $(n_i, s_i)$ and $(p_i, q_i)$ adopted in this example, and $(\hat{p}_i, \hat{q}_i)$ that minimizes $D_{MD}(P_{s_i,n_i})$

| $i$ | 1 | 2 | 3 | 4 | 5 | 6 | 7 | 8 | 9 | 10 |
|---|---|---|---|---|---|---|---|---|---|---|
| $n_i$ | 225 | 1125 | 1425 | 525 | 75 | 675 | 1275 | 975 | 825 | 375 |
| $s_i$ | 17 | 19 | 11 | 13 | 5 | 1 | 3 | 7 | 15 | 9 |
| $p_i$ | 0.1530 | 0.4440 | 0.9498 | 0.0512 | 0.6464 | 0.3531 | 0.7658 | 0.8359 | 0.2480 | 0.5528 |
| $q_i$ | 0.7553 | 0.2572 | 0.5560 | 0.1425 | 0.0571 | 0.9369 | 0.3530 | 0.8581 | 0.4413 | 0.6426 |
| $\hat{p}_i$ | 0.7658 | 0.9498 | 0.7658 | 0.7658 | 0.0512 | 0.6464 | 0.6464 | 0.6464 | 0.4440 | 0.4440 |
| $\hat{q}_i$ | 0.3530 | 0.5560 | 0.3530 | 0.3530 | 0.1425 | 0.0571 | 0.0571 | 0.0571 | 0.2572 | 0.2572 |

Table 4 Regression coefficients $\alpha_i$ and $\beta_i$, $i = 0, 1, \cdots, 5$

| $i$ | $\alpha_i$ | $\beta_i$ |
|---|---|---|
| 0 | 0.0503965655930105 | 0.1732384430740160 |
| 1 | 0.0009192090929939 | -0.0003419881219659 |
| 2 | -0.0034365336368717 | 0.0046060707097167 |
| 3 | -0.0000002204260424 | 0.0000001517932891 |
| 4 | -0.0000308280100528 | 0.0000140768595897 |
| 5 | 0.0022834280933421 | 0.0004673041886360 |

Treating $q$ and $p$ as the functions of $n$ and $s$, we can establish the following regression model

$$p = \alpha_0 + \alpha_1 n + \alpha_2 s + \alpha_3 n^2 + \alpha_4 ns + \alpha_5 s^2$$
$$q = \beta_0 + \beta_1 n + \beta_2 s + \beta_3 n^2 + \beta_4 ns + \beta_5 s^2 \quad . \tag{48}$$

Based on $(n_i, s_i)$ and $(\hat{p}_i, \hat{q}_i)$ in Table 3, $\alpha_i$ and $\beta_i$ are finally obtained, as shown in Table 4.

**5.2.2 Compare four sequences studied in other papers**

Reference [16] optimized the scrambling permutations of the generalized Halton sequence to generate four sequences $P_{2,1500}$, $P_{3,1500}$, $P_{5,1500}$ and $P_{6,1500}$, and compared them with the sequences generated by different authors [15,17,18,34-37]. In this subsection, the proposed DEM is also adopted to generate these sequences. The Halton sequence is treated as the initial sequence, and $(q, p)$ is obtained based on Eq. (48). Comparisons of the sequences generated by the DEM with that generated in Refs. [15-18,34-37] are given in Table 5, and the last line is the discrepancies of the sequences generated by the proposed DEM.

Table 5 Modified L2-discrepancy and star L2-discrepancy of sequences generated by different methods

| Refs. | $P_{2,1500}$ | | $P_{3,1500}$ | | $P_{5,1500}$ | | $P_{6,1500}$ | |
|---|---|---|---|---|---|---|---|---|
| | $D_{MD}$ | $D_{L_2}$ | $D_{MD}$ | $D_{L_2}$ | $D_{MD}$ | $D_{L_2}$ | $D_{MD}$ | $D_{L_2}$ |
| [15] | 0.002960 | 0.001737 | 0.004259 | 0.001669 | 0.009855 | 0.001649 | 0.014821 | 0.001363 |
| [17] | 0.001423 | 0.001123 | 0.002450 | 0.001273 | 0.007055 | 0.001382 | 0.010937 | 0.001257 |
| [34] | 0.001354 | 0.001006 | 0.002366 | 0.001252 | 0.006399 | 0.001388 | 0.010193 | 0.001257 |
| [35] | 0.002960 | 0.001737 | 0.002850 | 0.001225 | 0.007311 | 0.001405 | 0.010786 | 0.001387 |
| [36] | 0.001388 | 0.001129 | 0.002444 | 0.001279 | 0.006371 | 0.001295 | 0.010029 | 0.001237 |
| [18] | 0.001719 | 0.001183 | 0.002588 | 0.001290 | 0.006397 | 0.001300 | 0.010000 | 0.001211 |
| [37] | 0.001362 | 0.001029 | 0.002272 | 0.001237 | 0.007515 | 0.001399 | 0.010073 | 0.001272 |
| [16] | 0.001020 | 0.000830 | 0.002008 | 0.001107 | **0.005787** | 0.001316 | **0.009203** | 0.001208 |
| $U_{q,p}$ | **0.000932** | **0.000652** | **0.001986** | **0.000965** | 0.006092 | **0.001262** | 0.009814 | **0.001177** |

According to Table 5, there are two cases in all the four cases, i.e., $P_{2,1500}$ and $P_{3,1500}$, that the modified L2-discrepancies of the sequences generated by the proposed DME are smaller than that generated by Ref. [16], and in the other two cases the proposed DME generates the sequences with a little larger modified L2-discrepancies. However, the star L2-discrepancies of the sequences generated by the proposed DEM is in all cases smaller than that generated by Ref. [16]. Compared with other references, both the modified L2-discrepancies and the star L2-discrepancies of the sequences by the proposed DEM are in all cases smaller. The comparisons confirm the assumption that the sequence with the minimum potential energy principle can indeed generate a sequence with low-discrepancy.

**5.2.3 Eight test functions**

In this subsection, we examine the performance of all the 127 sequences in terms of the errors of eight integrals. These eight integrals are

$$I_i = \int_{[0,1]^s} F_i(x_1, \cdots, x_s) \mathrm{d}x_1 \cdots \mathrm{d}x_s,$$

where

$$F_1 = \sqrt{\frac{12}{s}} \left( \sum_{j=1}^{s} x_j - \frac{s}{2} \right), \quad F_2 = \sqrt{\frac{45}{4s}} \left( \sum_{j=1}^{s} x_j^2 - \frac{s}{3} \right), \quad F_3 = \sqrt{\frac{18}{s}} \left( \sum_{j=1}^{s} \sqrt{x_j} - \frac{2}{3} s \right);$$

$$F_4 = \prod_{j=1}^{s} \text{sign}(x_j - 0.5), \quad F_5 = \prod_{j=1}^{s} \left( \sqrt{\frac{15e^{15} + 15}{13e^{15} + 17}} \frac{e^{30x_j - 15} - 1}{e^{30x_j - 15} + 1} \right),$$

$$F_6 = \prod_{j=1}^{s} \left[ -2.4\sqrt{7}(x_j - 0.5) + 8\sqrt{7}(x_j - 0.5)^3 \right], \quad F_7 = \prod_{j=1}^{s} \left[ 2\sqrt{3}(x_j - 0.5) \right], \tag{49}$$

$$F_8 = \sqrt{\frac{2}{s(s-1)} \sum_{i=1}^{s} \sum_{j=i+1}^{s} f_i f_j}, \quad f_i = \begin{cases} 1, & x_i < 1/6, \text{ or, } x_i > 4/6 \\ 0, & x_i = 1/6, \text{ or, } x_i = 4/6. \\ -1, & x_i > 1/6, \text{ or, } x_i < 4/6 \end{cases}$$

The above eight integrand functions were proposed in Ref. [38] to test the accuracy of different low-discrepancy sequences. The functions $F_1$ to $F_3$ are smooth and satisfy conditions for proper quasi-Monte Carlo integration; the functions $F_4$ to $F_7$ have a great number of function extremes; and the functions $F_4$ and $F_8$ have flat regions and discontinuities. For all these eight functions, the exact values of the integrals are zero. The proposed sequences and the HS, HSO, SS, HW, Halton sequences are used to evaluate the numerical values of these eight integrals. As there are 127 cases with different dimensions and different numbers of sample points, and in fact no method could generate the best sequences for all cases, the relative advantage ratio (RAR) are used to compare the performance of different methods. Take the comparison between the proposed method with the HS method in calculating the integral of $F_1$ as an example. Both methods generate 127 sequences respectively, and these sequences are used to calculate the integral of $F_1$. Therefore, both methods produce 127 pairs of numerical results which would be compared in turn. Among the 127 comparisons, the number of times that the HS result is more accurate is denoted as $n_{HS}$, and the ratio of $n_{HS}$ to 127 is called the RAR of the HS in calculating the integral of $F_1$. The RAR of the HS in the modified L2-discrepancy or the star L2-discrepancy can also be defined in a similar way. The RARs of different methods in calculating the integrals of $F_i$, the modified L2-discrepancy and the star L2-discrepancy are listed in Table 6.

Table 6 The relative advantage ratios of different methods in calculating the integrals of $F_i$, the modified L2-discrepancy and the star L2-discrepancy

|  | $F_1$ | $F_2$ | $F_3$ | $F_4$ | $F_5$ | $F_6$ | $F_7$ | $F_8$ | $D_{MD}$ | $D_{L_2D}$ |
|---|---|---|---|---|---|---|---|---|---|---|
| HS | 22.0% | 26.0% | 37.0% | 42.5% | 51.2% | 52.0% | 49.6% | 29.1% | 3.1% | 24.4% |
| HSO | 22.8% | 15.7% | 29.1% | 52.8% | 56.7% | 63.0% | 67.7% | 36.2% | 10.2% | 18.9% |
| SS | 1.6% | 2.4% | 8.7% | 50.4% | 44.9% | 43.3% | 48.0% | 27.6% | 0.8% | 22.0% |
| HW | 4.7% | 5.5% | 5.5% | 40.9% | 50.4% | 58.3% | 48.8% | 32.3% | 1.6% | 23.6% |
| Halton | 0.0% | 0.0% | 0.0% | 56.7% | 61.4% | 45.7% | 58.3% | 28.3% | 0.0% | 0.0% |

It can be seen from Table 6 that for the functions $F_1$, $F_2$, $F_3$ and $F_8$, the RAR of all five methods is less than 40%, indicating that the accuracy of most of the 127 sequences generated by the proposed DEM is the best in calculating these four types of functions. For the functions $F_4$ to $F_7$, the RAR values of all these methods jump around 50%, with a minimum of 40.9% and a maximum of 67.7%, which indicates that the accuracy of the sequences generated by all the several methods is similar in calculating these four types of functions.

In order to compare more clearly the performance of each sequence, we introduce the test function discrimination coefficient $b_1$ and the discrepancy discrimination coefficient $b_2$ as follows

$$b_1(P_{s,n}) = \text{sum}\left[\text{sign}(|F_1|-|F_2|)\right], \quad b_2(P_{s,n}) = \text{sum}\left[\text{sign}(|D_1|-|D_2|)\right]. \tag{50}$$

where $F_i$ is the vector containing the numerical results of all the eight test functions computed by different methods, $D_i$ is the vector containing the modified L2-discrepancy and the star L2-discrepancy of the sequences generated by different methods, and "sum$[b]$" represents the sum of all the elements in the vector $b$. Take the comparison between the HSO method and the proposed DEM as an example: $F_1 = \{I_{1,\text{HSO}}, \cdots, I_{8,\text{HSO}}\}$ is the vector containing the numerical results of all the eight test functions computed by the HSO sequences, and $F_2 = \{I_{1,\text{DEM}}, \cdots, I_{8,\text{DEM}}\}$ is the vector containing the numerical results of all the eight test functions computed by the proposed DEM sequences. $D_1 = \{D_{\text{MD,HSO}}, D_{L_2,\text{HSO}}\}$ is the discrepancy vector of the HSO sequence; and $D_2 = \{D_{\text{MD,DEM}}, D_{L_2,\text{DEM}}\}$ is the discrepancy vector of the sequence generated by the proposed DEM. If $b_1(P_{s,n}) > 0$, it means that among the eight tests, the number of times that the HSO sequence is more accurate is larger than that of the DEM sequences, and the converse is also true. The same is true for the discrepancy discrimination. The test function comparisons of the proposed DEM sequences with the HSO sequences and the HS sequences are drawn in Fig. 6, and the discrepancy comparisons of these sequences are drawn in Fig. 7. It can be seen from Figs. 6 and 7 that the sequences generated by the proposed DEM perform far better than that generated by the HSO or the HS method, in terms of both the test function comparison and the discrepancy comparison.

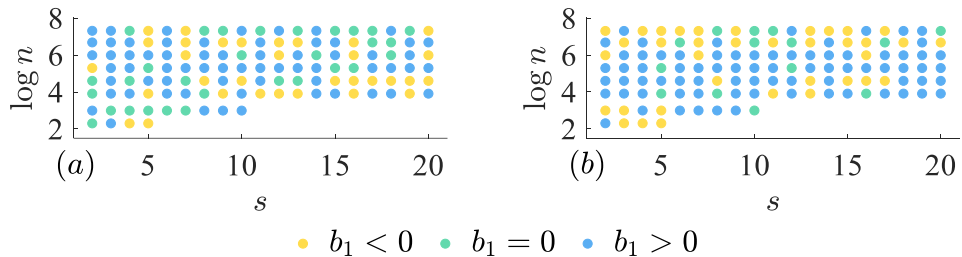

Figure 6 Test function comparison: (a) DEM with $U_{q,p}$ vs HSO; (b) DEM with $U_{q,p}$ vs HS

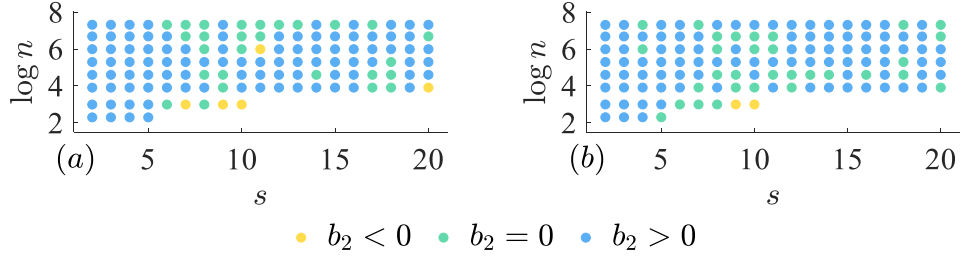

Figure 7 Discrepancy comparison: (a) DEM with $U_{q,p}$ vs HSO; (b) DEM with $U_{q,p}$ vs HS

## 5.3 Practical stochastic simulation problems

### 5.3.1 Kraichnan-Orszag equation

Kraichnan-Orszag equation was first proposed by Kraichnan [39] to analyze the evolution of the inviscid and interacting shear waves in the turbulent flow, and then was employed by Orszag [40] to study the dynamical properties of the truncated Wiener-Hermit expansions in the theory of turbulence. It can be written as

$$\frac{du_1}{dt} = u_2 u_3; \quad \frac{du_2}{dt} = u_3 u_1; \quad \frac{du_3}{dt} = -2u_1 u_2, \tag{51}$$

with the random initial condition

$$u_1(0) = 1 + 0.1\xi_1; \quad u_2(0) = 2 + 0.2\xi_2; \quad u_3(0) = 1 + 0.1\xi_3, \tag{52}$$

where $\xi_i \in [-1,1]$ are independent and uniformly distributed random variables. In this example, we use the low-discrepancy sequences to numerically obtain the means and standard derivations (SDs) of the random responses $u_i$. The Euler midpoint scheme with the time step $\Delta t = 0.1\,\text{s}$ is used for the time integration.

The proposed DEM with $U_{q,p}$, HSO, HS, HW, and SS sequences are used to solve this problem, and the number of samples is $n = 200$. The random sampling sequences with $10^7$ sample points is used to produce the reference solution. The relative errors of numerical results computed by using different sequences are listed in Table 7, and the relative error is defined as:

$$\varepsilon_u = \frac{\|u - u_r\|_2}{\|u_r\|_2}, \quad \|u_r\|_2 = \sqrt{\int_0^{20} u_r^2(t)\,dt}, \quad \|u - u_r\|_2 = \sqrt{\int_0^{20} \left[u_r(t) - u(t)\right]^2 dt}\,. \tag{53}$$

where $u$ represents the solution calculated by different sequences and $u_r$ represents the reference solution. The means and SDs of $u_1$ computed based on the sequence generated by the DEM with $U_{q,p}$ and the reference solutions are plotted in Fig. 8. It can be seen from Table 7 and Fig. 8 that the $U_{q,p}$ solutions are in good agreement with the reference solution, and the relative errors of the $U_{q,p}$ solutions are smaller than that of the other low-discrepancy sequences.

Table 7 Relative errors of means and SDs calculated by different sequences

| | Relative errors of means | | | | Relative errors of SDs | | | | |
|---|---|---|---|---|---|---|---|---|---|
| | $U_{q,p}$ | HSO | HS | HW | SS | $U_{q,p}$ | HSO | HS | HW | SS |
| $u_1$ | 0.44% | 0.76% | 0.48% | 0.68% | 6.67% | 0.22% | 0.38% | 0.36% | 0.75% | 2.70% |
| $u_2$ | 0.03% | 0.04% | 0.05% | 0.10% | 0.40% | 0.44% | 0.69% | 0.72% | 0.68% | 2.28% |
| $u_3$ | 0.44% | 0.73% | 0.46% | 0.74% | 6.69% | 0.22% | 0.36% | 0.35% | 0.72% | 2.72% |

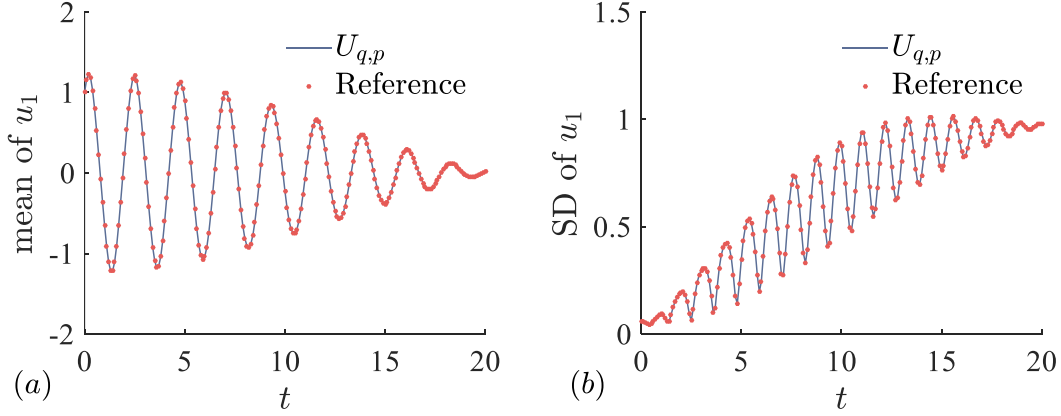

Figure 8 Comparison of the $U_{q,p}$ solution and the reference solution: (a) mean of $u_1$; (b) SD of $u_1$

### 5.3.2 Plane steel frame

As the second example, a 20-storey floor-shear structure [41] with random story stiffness is taken into account. All the stories have equal masses, $m = 168750 \text{kg}$, and the story stiffness for each story is random, i.e.,

$$k_i = 2.72 \times 10^8 (1+\varepsilon_i) \text{ N/m}, \quad i = 1, 2, \cdots, 20, \tag{54}$$

where $\varepsilon_i \in [-0.3, 0.3]$ are independent and uniformly distributed random variables. The means and SDs of the natural vibration frequencies of the frame are obtained by using the HS, HSO, HW and proposed sequences, and the number of samples is $n = 400$. The solutions obtained using the MCS with $10^7$ random samples are considered as the reference solutions. The relative errors $e_{\max}$ of the frequencies are defined by

$$e_{\max} = \max_{1 \le i \le 5} \left| \frac{y_i - y_{i,\text{MCS}}}{y_{i,\text{MCS}}} \right|, \tag{55}$$

in which $y_{i,\text{MCS}}$ is the mean or SD of the $i$-th frequency computed using the MCS with $10^7$ random samples, and $y_i$ is the mean or SD of the $i$-th frequency computed using different low-discrepancy sequences. The means and SDs computed using different methods and the corresponding relative errors are listed in Table 8. It can be observed from Table 8 that the means and SDs computed using the sequences generated by the proposed DEM with $U_{q,p}$ are far more accurate than that computed using other low-discrepancy sequences.

Table 8 Means and SDs computed using different methods and the corresponding relative errors

| $i$-th | mean | | | | | SD | | | | |
|---|---|---|---|---|---|---|---|---|---|---|
| | MCS | $U_{q,p}$ | HSO | HS | HW | MCS | $U_{q,p}$ | HSO | HS | HW |
| 1 | 2.9413 | 2.9415 | 2.9407 | 2.9420 | 2.9411 | 0.0166 | 0.0168 | 0.0156 | 0.0167 | 0.0160 |
| 2 | 8.8072 | 8.8098 | 8.8083 | 8.8047 | 8.8110 | 0.1486 | 0.1491 | 0.1396 | 0.1449 | 0.1553 |
| 3 | 14.6232 | 14.6213 | 14.6140 | 14.6148 | 14.6377 | 0.4106 | 0.4131 | 0.3741 | 0.4086 | 0.3837 |
| 4 | 20.3563 | 20.3647 | 20.3607 | 20.3383 | 20.3300 | 0.7963 | 0.8026 | 0.7818 | 0.7663 | 0.8658 |
| 5 | 25.9747 | 25.9729 | 25.9962 | 25.9841 | 25.9766 | 1.2991 | 1.3108 | 1.2726 | 1.1825 | 1.4043 |
| $e_{\max}$ | - | 0.041% | 0.083% | 0.089% | 0.129% | - | 1.246% | 8.885% | 8.973% | 8.732% |

## 5.4 Sequence uniformly scattered in non-cube

One advantage of the proposed DEM is that it is easy to generate the uniformly distributed sequence in the non-cube. In this section, some examples about the sequence uniformly scattered in the non-cube are given. Let the non-cube considered be $\Omega := \{x \in \mathbb{R}^s, \ g(x) \leq 0\}$. The analysis is still based on the proposed DEM, but the periodic boundary model shown in subsection 4.2 can not be used to deal with the points moving out the bound of non-cube. Considering the gravity model consisting of $n$ stars which are scattered evenly in the non-cube $\Omega$, the size of the influence range (i.e., the Voronoi cell) of each star is almost the same, and hence the radius of the inscribed circle in each Voronoi cell is also close to the same. Based on this observation, the distance from each point to the boundary is also taken into account, and the potential energy can be modified as

$$U_{\mathrm{E},\infty,g} = G\frac{1}{d_{\min}}, \quad d_{i,g} := \min_{g(x)=0} \sqrt{(x_i - x)^{\mathrm{T}}(x_i - x)}, \quad d_{\min} := \min\left\{\min_{1 \leq i < j \leq n} d_{\mathrm{E},ij}^p, \min_{1 \leq i \leq n}(2d_{i,g})\right\}, \quad (56)$$

where $d_{\mathrm{E},ij}$ represents the Euclidean distance between points $x_i$ and $x_j$, as shown in Eq. (8), and $d_{i,g}$ represents the distance from the $i$-th point to boundary $g(x) = 0$. Therefore, the load in the Eq. (20) is modified as follows:

$$f_{ik} = \begin{cases} -Gd_{\min}^{-3} a_{ijk}, & \text{if } d_{\mathrm{E},ij} = d_{\min} \\ -4Gd_{\min}^{-3}(x_{ik} - x_{ik,g}), & \text{if } d_{i,g} = 0.5d_{\min} \\ 0, & \text{otherwise} \end{cases} \quad (57)$$

where $x_{ik,g}$ represents the $k$-th coordinate value of the point on the boundary that are more closer to $x_i$ than any other points on the boundary, and the expression of $a_{ijk}$ is shown in Eq. (15). Taking the two-dimensional case as an example, we study the regions described by the following functions,

$$\begin{aligned} g_1(x_1, x_2) &= 17x_1^2 - 16|x_1|x_2 + 17x_2^2 - 225, \quad g_2(x_1, x_2) = x_1^2|x_1| + 10x_2^2 - 100, \\ g_3(x_1, x_2) &= y^2 + 2|x| - 20, \quad g_4(x_1, x_2) = |x| + 2|y| - 5. \end{aligned} \quad (58)$$

10000 random point are generated, and the first 100 points satisfying $g_i(x_1,x_2)\leq 0$ are selected as the initial sequence. Then the DEM with the potential energy Eq. (56) is used to generate the points that are uniformly distributed in the non-square region. The comparisons between the sequences obtained by the DEM and the initial random sequences are shown in the Fig. 9. Because there is no corresponding discrepancy expression to define the uniformity of the sequence distributed in the non-cube region, we use the potential energy $U_{E,\infty,g_i}$ to measure the uniformities of the random sequence and the sequence calculated by the proposed DEM.

As can be seen from Fig. 9, the optimization effect of the DEM is very significant, and the points generated by the DEM are scattered uniformly in all the four non-cubes. Using the potential energy $U_{E,\infty,g_i}$ to compare the uniformity of the DEM sequence with that of the initial random sequence, the minimum optimization ratio is 80.1%, and the maximum optimization ratio 89.3%, which are corresponding to $g_1$ and $g_4$, respectively. The comparisons show clearly that the proposed DEM can still produce a very uniform sequence for the non-cube.

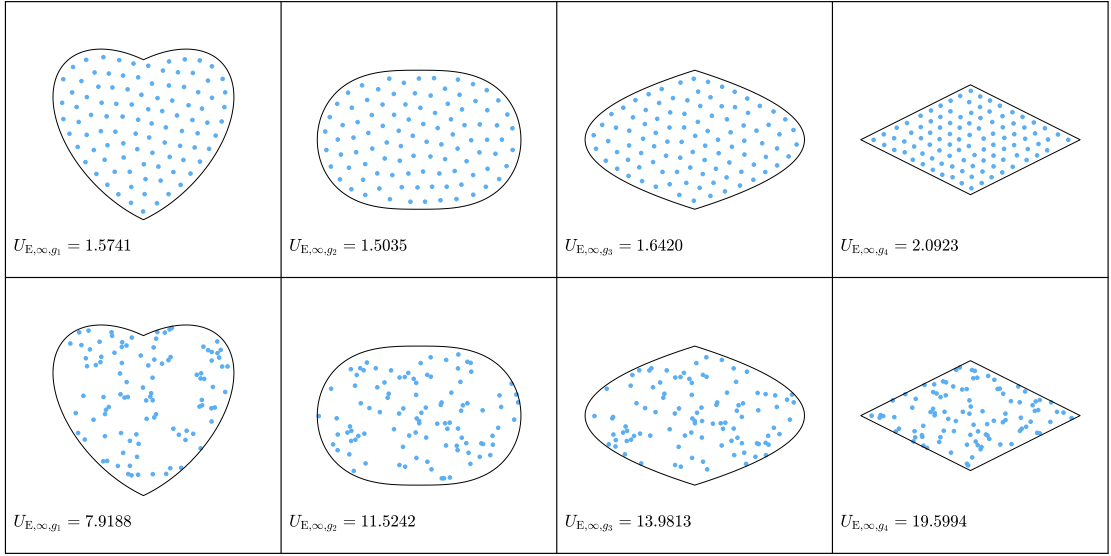

Figure 9 Comparisons between the sequences generated by the DEM and the random initial sequences, $n=100$. From left to right are the regions described by $g_1(x_1,x_2)$, $g_2(x_1,x_2)$, $g_3(x_1,x_2)$ and $g_4(x_1,x_2)$ functions respectively. The first line displays the sequences generated by the DEM, and the second line the random sequences.

# 6. Conclusion

Low-discrepancy sequence and multi-bodies problem are two different types of issues. Low-discrepancy sequence is a foundational and important tool in many fields, such as quantum physic and artificial intelligence. How to construct the low-discrepancy has always been an important issue concerned by statisticians. On the other hand, the multi-bodies problem is a classical and difficult problem in physics. Physicists and mathematicians have been trying to find the dynamical solutions to multi-bodies problem for a long time, and the static solution did not seems to get much attention. Although there are some scholars who tried to use the low-discrepancy sequence to obtain the dynamical solutions to the multi-bodies problems, it seems that there is no much connection between these two issues. The static state of multi-bodies with the same mass means the resultant force acted on each body is zero, and the potential energy of the multi-bodies system is the smallest. Considering the form of the force acted on each body is the same, the bodies in the static state are scattered uniformly. This phenomenon motives the author to think of the relationship between the static solution to multi-bodies and low-discrepancy sequence, and the relationship between the potential energy and the discrepancy.

To obtain the static solutions of the multi-bodies with the same mass, a dynamical evolutionary model (DEM) based on the Hamilton variational principle is proposed. In the DEM, the forces acted on each body is derived in terms of two different types of potential energies. To construct these two potential energies, two forms of distances are discussed, and a new distance based on the $q$-norm is proposed. The proposed DEM also contains an artificial damping that can dissipate the total energy of the multi-bodies and lead to the static solution. The central difference scheme is adopted to transfer the DEM to the discrete form which is an iterative scheme. The selections about the mass and the damping coefficient are discussed to ensure the convergence of the iterative scheme. It has been verified that the iterative scheme will converge to the static solution to the multi-bodies problem. In addition, the restarting technique is proposed to keep the unbiasedness of solutions during the iterative process.

A series of numerical examples are given to study the proposed DEM of the multi-bodies problem in high-dimensional space. In the first group of examples, the relationship between the potential energy and the discrepancy are discussed by using 11 DEMs with different numbers of samples and different dimensions. The numerical results show clearly that there is a significance positive correlation between the potential energy and the discrepancy during the evolutionary iterative process, which means the potential energy can be seen as an index to measure the uniformity of the sequences, and more importantly, the static solution to the multi-bodies problem is a low-discrepancy sequence in the high-dimensional space.

Based on the proposed DEM, we have generated a series of low-discrepancy sequences $P_{s,n}$ with $s \in \{1, 2, \cdots, 20\}$, $n \in \{10, 20, 50, 100, 200, 400, 800, 1500\}$. All these sequences can be downloaded from https://rocewea.com/2.html. Comparisons between these generated sequences and other low-discrepancy sequences are given by using eight test functions and two practical applications in the stochastic dynamical problems. In all the comparisons, the sequences generated using the proposed DEM perform better than other low-discrepancy sequences. The ability of the proposed DEM to generate the points uniformly distributed in the non-cube is also tested by four examples. Through the theoretical analyses and numerical examples, it can be concluded the proposed DEM can generate the unbiased low-discrepancy sequences in the high-dimensional cube or non-cube.

In future, we will generate more sequences with larger dimensions and more points in the high-dimensional cube. We will also generate more sequences in some typical non-cubes, such as sphere and ellipsoid. All the generated sequences will be uploaded to https://rocewea.com/2.html.

# Acknowledgements

The authors are grateful for the support of the Natural Science Foundation of China (Nos. 11472076, 51609034), Dalian Youth Science and Technology Star Project (No. 2048RQ06), and Fundamental Research Funds for the Central Universities (No. DUT20RC(5)009).